\documentclass[final,1p,times]{elsarticle}

\usepackage{amssymb}

\usepackage[figuresright]{rotating}
\usepackage{amsmath}
\usepackage{amscd}
\usepackage{wrapfig}
\usepackage{blkarray}
\usepackage{hyperref}
\usepackage{color}

\usepackage{amsfonts}
\usepackage{mathrsfs}
\usepackage{stmaryrd}
\usepackage{mathtools} 
\usepackage{amsfonts} 

\usepackage{algorithm}
\usepackage{algorithmic}
\usepackage{multirow}

\usepackage{dsfont}
\usepackage{url}

\newtheorem{example}{Example}[section]

\newtheorem{definition}{Definition}[section]
\newtheorem{theorem}{Theorem}[section]

\newtheorem{corollary}[theorem]{Corollary}
\newtheorem{lemma}[theorem]{Lemma}
\newtheorem{remark}[theorem]{Remark}

\begin{document}

\makeatletter
  \newcommand\figcaption{\def\@captype{figure}\caption}
  \newcommand\tabcaption{\def\@captype{table}\caption}
\makeatother

\newcommand{\abs}[1]{\lvert#1\rvert}

\begin{frontmatter}



\title{Index reduction of differential algebraic equations by differential algebraic elimination}


\author[Casit,LiU]{Xiaolin Qin \corref{cor1}}
\ead{qinxl@casit.ac.cn}
\author[Casit]{Lu Yang}
\author[Casit]{Yong Feng}
\author[Buas]{Bernhard Bachmann}
\author[LiU]{Peter Fritzson}

\cortext[cor1]{Corresponding author}
\address[Casit]{Chengdu Institute of Computer Applications, Chinese Academy of Sciences, Chengdu 610041, China}
\address[LiU]{Department of Computer and Information Science, Link\"{o}ping University, Link\"{o}ping SE-581 83, Sweden}
\address[Buas]{Department of Mathematics and Engineering, Bielefeld University of Applied Sciences, Bielefeld D-33609, Germany}

\begin{abstract}
High index differential algebraic equations (DAEs) are ordinary differential equations (ODEs) with constraints and arise frequently from many mathematical models of physical phenomenons and engineering fields. In this paper, we generalize the idea of differential elimination with Dixon resultant to polynomially nonlinear DAEs. We propose a new algorithm for index reduction of DAEs and establish the notion of differential algebraic elimination, which can provide the differential algebraic resultant of the enlarged system of original equations. To make use of structure of DAEs, variable pencil technique is given to determine the termination of differentiation. Moreover, we also provide a heuristics method for removing the extraneous factors from differential algebraic resultant. The experimentation shows that the proposed algorithm outperforms existing ones for many examples taken from the literature.

\end{abstract}


\end{frontmatter}


\section{Introduction}
\label{intro}
Modeling with differential algebraic equations (DAEs) plays a vital role in a variety of applications \cite{KM2006}, for constrained mechanical systems, control theory, electrical circuits and chemical reaction kinetics, and many other areas. In general, it is directly numerical computations difficult to solve the system of DAEs. The index of DAEs is  a measure of the number of times needed to differentiate it to get its equivalent low index or ordinary differential equations (ODEs). There exist many different index concepts for the specific DAEs, such as the differentiation index \cite{BCP1996, CG1995}, perturbation index \cite{CG1995, HW1996}, tractability index \cite{M1992}, structural index \cite{P2001}, and Kronecker index \cite{TI2008}. There has been considerable research for the general linear and low index DAEs \cite{KM2006, M1992, P1988, TI2008}. In particular, it may only solve some special DAEs by the directly numerical methods \cite{EH2009, L2014}. It is more difficult to solve the system of high index nonlinear DAEs \cite{BCP1996, C1993, CG1995, G1988, MS1993, P2001}.  Therefore, index reduction techniques may be necessary to get a solution \cite{BCP1996}. 

Index reduction in the pre-analysis of DAEs solving  is an active technique of research. It is equivalent to applying a sequence of differentiations and eliminations to an input system of DAEs. In \cite{P1988}, Pantelides gave a systematic way to reduce the high index DAEs to low index one, by selectively adding differentiated forms of the equations already appear in the system. However, the algorithm can succeed yet not correctly in some instances and be just first order \cite{RMB2000} . Campbell's derivative array theory needs to be computationally expensive especially for computing the singular value decomposition of the Jacobian of the derivative array equations using nonlinear singular least squares methods \cite{C1993}. In \cite{MS1993}, Mattsson et al. proposed the dummy derivative method based on Pantelides' algorithm for index reduction, which is an algebraic viewpoint. In \cite{P2001}, Pryce proposed the signature matrix method (also called $\Sigma$-method), which can be viewed as an extension of Pantelides' method for any order. Recently, Wu et al. generalized the $\Sigma$-method for DAEs to partial differential algebraic equations with constraints (PDAEs) \cite{WRI2009}. Qin et al. presented the structural analysis of high index DAEs for process simulation by the $\Sigma$-method \cite{QWFR2013}. But the $\Sigma$-method relies heavily on square (i.e. the same number of DAEs and dependent variables) and sparsity structure, which is confronted with the same drawback that can succeed yet not correctly in some DAEs arising from the specific types. 

A principal aim of this paper is the development of an efficient differential elimination approach for index reduction of DAEs that extends the direct elimination treatment of \cite{YZZ2012}. From the algebraic standpoint, differential elimination algorithms which are key for simplifying systems of polynomially differential equations and computing formal power series solutions for them. The underlying theory is the differential algebra of Ritt \cite{R1950} and Kolchin \cite{K1973}. Differential elimination algorithm in algebraic elimination theory is an active field and powerful tools with many important applications \cite{F1997, GLY2013, RLW2001, R2013, YZZ2012,  ZYG2014}. Almost all of the authors focus on the differential elimination theory for ODEs. Only Reid et al. presented an effective algorithm for computing the index of polynomially nonlinear DAE and a framework for the algorithmic analysis of perturbed system of PDAEs. This underlies the jet space approach based on differential geometry. 

In this paper, we want to promote the efficient differential elimination algorithm as natural generalization of DAEs, which is a direct and elementary approach. In particular, differential elimination with Dixon resultant can be solved by eliminating serval variables at a time, simplifying the system with respect to its constraints, or determining its singular cases \cite{YZZ2012}. We can directly transform the system of DAEs to its equivalent ODEs by differential algebraic elimination. Differential algebraic elimination is to apply a finite number of differentiations and eliminations to uncover all hidden constraints of system of DAEs. We define the new minimum differentiation time, which is the weak differentiation index for DAEs/ODEs. It can be used as a unified formulation of differentiation times for differential elimination of ODEs and differential algebraic elimination of DAEs. Meanwhile, we provide the new index reduction with variable pencil and the notion of differential algebraic resultant. In order to overcome the drawback of factoring a large polynomial system \cite{YZZ2012}, we consider a heuristics method for removing the extraneous factors from the differential algebraic elimination matrix. Our algorithm is also suitable for the non-square nonlinear DAEs/ODEs. To the best of our knowledge, it is the first time that the generalized Dixon resultant formulation has been directly extended to the system of DAEs.

The rest of the paper is organized as follows. Section 2 gives a brief description of the generalized Dixon resultant formulation, and analyzes the size of Dixon matrix and the complexity of computing the entries of Dixon matrix. Section 3 proposes the new index reduction procedure for the system of DAEs and defines the weak differentiation index. Section 4 provides the differential algebraic elimination algorithm and some basic properties of differential algebraic resultant. Section 5 presents some
specific examples in detail and comparisons of our algorithm for the system of ODEs. The final section concludes this paper.

\section{Generalized Dixon resultant formulation}
\label{genDixon}
Following Kapur et al. \cite{CK2003, CK2004, KS1996, KSY1994, ZF2005, ZF2009}, we introduce the concept of generalized Dixon resultant formulation and its properties. This technique will play a central role in our subsequent analysis. Let $X=\{x_1, x_2, \cdots, x_n\}$ and $\bar{X}=\{\bar{x}_1, \bar{x}_2, \cdots, \bar{x}_n\}$ be two sets of $n$ variables, respectively. The determinant of a square matrix $A$ is denoted by $det(A)$.
\begin{definition}
Let $\mathcal{F}=\{f_1, f_2, \cdots, f_{n+1}\} \subset \mathbb{Q}[X]$ be a set of $n+1$ polynomials in $n$ variables. The cancellation matrix $\mathcal{C}_\mathcal{F}$ of $\mathcal{F}$ is the $(n+1)\times(n+1)$ matrix as follows:
 \begin{equation*}
\begin{array}{c}
\mathcal{C}_\mathcal{F} = \begin{bmatrix}
f_1(x_1, x_2, \cdots, x_n) & \cdots & f_{n+1}(x_1, x_2, \cdots, x_n) \\
f_1(\bar{x}_1, x_2, \cdots, x_n) & \cdots & f_{n+1}(\bar{x}_1, x_2, \cdots, x_n) \\
f_1(\bar{x}_1, \bar{x}_2, \cdots, x_n) & \cdots & f_{n+1}(\bar{x}_1, \bar{x}_2, \cdots, x_n) \\
\vdots& \vdots & \vdots \\
f_1(\bar{x}_1, \bar{x}_2, \cdots, \bar{x}_n) & \cdots & f_{n+1}(\bar{x}_1, \bar{x}_2, \cdots, \bar{x}_n)
\end{bmatrix},
\end{array}
\end{equation*}
where $f_i(\bar{x}_1, \bar{x}_2, \cdots, \bar{x}_k, x_{k+1}, x_{k+2}, \cdots, x_n)$ stands for uniformly replacing $x_j$ by $\bar{x}_j$ for all $1 \leq j \leq k \leq n$ in $f_i$. The Dixon polynomial of $\mathcal{F}$ is denoted by $\theta_\mathcal{F} \in \mathbb{Q}[X, \bar{X}]$,
\begin{equation}
\theta_\mathcal{F} =\frac{det(\mathcal{C}_\mathcal{F})}{\prod_{i=1}^{n}(x_i-\bar{x}_i)},
\end{equation}
the row vector of Dixon derived polynomials of $\mathcal{F}$ is denoted by $P_\mathcal{F}$, and Dixon matrix of $\mathcal{F}$ is denoted by $D_\mathcal{F}$ as follows,
\begin{equation}
\theta_\mathcal{F} =P_\mathcal{F} V_{\bar{X}}(\theta_\mathcal{F}) =V_{X}(\theta_\mathcal{F}) D_\mathcal{F}  V_{\bar{X}}(\theta_\mathcal{F}),
\end{equation}
where $V_{\bar{X}}(\theta_\mathcal{F})$ is a column vector of all monomials in $\bar{X}$ which appears in $\theta_\mathcal{F}$, and $V_{X}(\theta_\mathcal{F})$ is a row vector of all monomials in $X$ which appears in $\theta_\mathcal{F}$. The determinant of $D_\mathcal{F}$ is called the Dixon resultant, denoted by $res(f_1, f_2, \cdots, f_{n+1})$.
\end{definition}
It is well known that Dixon resultant is a projection operator whose vanishing is a necessary condition for the system $\mathcal{F}$ to have a common affine solution. However, the Dixon matrix may be non-square then its determinant cannot be directly computed. Even if it is square, the Dixon resultant vanishes identically without providing any information for the affine solutions. In \cite{KSY1994}, Kapur et al. presented a heuristic method to remedy the drawback by extracting a non-trivial projection operator.
\begin{lemma}\label{lem:KSY}(\cite{KSY1994})
If there exists a column which is linearly independent of all other columns in $D_\mathcal{F}$, then the determinant of any non-singular rank submatrix of $D_\mathcal{F}$ is a non-trivial projection operator.
\end{lemma}
\begin{remark}
From Lemma \ref{lem:KSY}, this method may fail if there is no column which is linearly independent of all other columns in $D_\mathcal{F}$. However, the method is quite efficient and practical as demonstrated in \cite{KSY1994, YZZ2012, ZF2005, ZF2009}, and such failure is very rare even never occurred on the numerous problems. Furthermore, the projection operator may contain extraneous factors in the Dixon resultant.
\end{remark}
In this article, we shall use the following properties of Dixon resultant.

\begin{lemma}\label{lem:lincomb}(\cite{ZF2009})
Dixon resultant can be expressed as a linear combination of original polynomial system $\mathcal{F}$,
\begin{equation}
res(f_1, f_2, \cdots, f_{n+1}) = \sum_{i=1}^{n+1} K_i f_i,
\end{equation}
where $K_i$ is a polynomial with respect to $X$ and can be deduced from $P_\mathcal{F}$. Moreover, it has been proved that the extraneous factors mentioned above may include three parts which are taken from $P_\mathcal{F}$, $D_\mathcal{F}$ and the resulting resultant expression by substituting $P_\mathcal{F}$, respectively.
\end{lemma}
\begin{remark}
Extraneous factors arising from Dixon resultant is a troublesome problem when it is used for elimination in a variety of applications. Gather-and-Sift method \cite{YH1995} is a complete method to remove extraneous factors by the simplicial decomposition algorithm. But it suffers from very high computational complexity because of the intermediate expression swell in symbolic computation. Therefore, we mainly use the technique based on Lemma \ref{lem:lincomb}, which can be viewed as a heuristic method.
\end{remark}

\begin{theorem}\label{thm:dixsizcom}
The size of Dixon matrix is at most $n!\prod_{i=1}^{n}d_i \times n!\prod_{i=1}^{n}d_i$, and the complexity of computing the entries of Dixon matrix is $\mathcal{O}(d_{1}^2(n!\prod_{i=2}^{n}d_i)^3)$ in the worst case, where $d_i$ is the highest degree of variable $x_i$.
\end{theorem}
\begin{proof}
Similar to the proof of computing the entries of Dixon matrix in the bivariate case and combine with the multivariate Sylvester resultant and the general case in \cite{ZF2005, ZF2010}. $\hfill \square$
\end{proof}
\begin{remark}
Here we give the size and computational complexity of Dixon matrix in the general setting. In particular, the complexity of computing the entries of Dixon matrix is a new result. The highest degree $d_i$ of variable $x_i$ can be obtained by using the algorithm in \cite{QSLF2014}. To make use of sparsity in polynomial systems, bound on the size of Dixon matrix of the specific systems is derived in terms of their Newton polytopes in \cite{KS1996}.
\end{remark}

\section{Index reduction algorithm}
\label{index}
In this section, let $\delta=d/dt$ denote the differentiation operator, let $R$ be a differential ring, i. e., a commutative ring with unit and a differentiation $\delta$ acting on it. Let $\mathbb{N}_0 = \{0, 1, \cdots, n, \cdots\}$, $\upsilon \in \mathbb{N}_0^n$ represents a multi-index $\mathbb{\upsilon} = (\upsilon_1, \upsilon_2, \cdots, \upsilon_n)^T\footnote{where $^T$ denotes the transposition, which is the same way for the rest of this article.}$, $m \in \mathbb{N}, \mathbb{Y} = \{y_1, y_2, \dots, y_m \}$. If $r\in \mathbb{N}_0$, then order of $\delta^r$ is $ord(\delta^r) = r$, we denote $y_{j}^{(k)}$ the $k$-th derivative of $y_j$ and $y_j^{[k]}$ to represent the set $\{ y_j^{(i)}, i=1, 2, \cdots, k\}$, in particular, $y_{j}^{(1)}$ and $y_{j}^{(2)}$ denote $\dot{y}_j$ and  $\ddot{y}_j$  in the following examples for notational simplicity. $|\mathbb{Y}| = m$, $|\cdot|$ denotes the cardinality of a set.

We give a new index reduction technique for DAEs and define the weak differentiation index. With loss of generality, consider $n$ polynomially DAEs with $m$ dependent variables $y_j  = y_j (t) $ with $t$ a scalar independent variable, of the form
\begin{equation}\label{eqn:original}
f_i = f(t, \ the \ y_j \ and \ derivatives \ of \ them ) = 0,  \ 1 \leq i \leq n, \ 1 \leq j \leq m.
\end{equation}
It is the following equivalent form from the above notations,
\begin{equation}\label{eqn:equivorignial}
f_i =c_{i0} + \sum_{k=1}^{l_i} c_{i k}\mathbb{P}_{ik}.
\end{equation}
where $c_{i0}, c_{ik}$ are the coefficients that are the known forcing functions with $t$ or the constants, $\mathbb{P}_{ik}= (\mathbb{Y}^{[r_j]})^{\alpha_{ik}}$ is a monomial in $\{y_1, y_2, \cdots, y_m, y_1^{(1)}, \cdots, y_1^{(r_1)}, \cdots, y_m^{(1)}, \cdots, y_m^{(r_m)}\}$ with exponent vector $\alpha_{ik}$ and $l_i = |\alpha_{ik}| \geq 1 $. In particular, the highest degree of $y_j$ and its derivative $y_{j}^{(r_j)}$ denote $d_j$ and $d_{jr_j}$ in $\{f_1, f_2, \cdots, f_{n}\}$, respectively.

In order to compute the differential algebraic resultant in Section \ref{diffRes}, the outline of index reduction procedure is as follows:
\begin{description}
  \item[Phase 1] Initialization. \\
    (a) Collect every dependent variable $y_j$ and its derivative $y_j^{[r_j]}$ for each $f_i$, and then gather the set of dependent variables $\mathbb{V}$.\\
    (b) Sort $\mathbb{V}$ for every $y_j$ and $y_j^{[r_j]}$ into a lexicographic order under assumption of ordering $\cdots \succ y_2 \succ y_1^{(r_1)} \succ \cdots \succ y_1^{(1)} \succ y_1$, and $\mathbb{V}_j$ represents the set of $y_j$ and its derivative $y_j^{[r_j]}$.\\
    (c) Construct a matrix $\mathcal{M}=(m_{ij})$, which is called \emph{variable pencil}, defined for (\ref{eqn:original}) by
    \begin{equation}\label{varpencil}
        m_{ij}=\begin{cases}
        1 \ \  the \ y_j\ or \ its \ derivative\ y_j^{[r_j]}\ appears\ in\ equation\ f_i, \\
        0\ \ if \ the\ variable\ does\ not\ occur,
\end{cases}
    \end{equation}
    where $row(\mathcal{M})$ and $col(\mathcal{M})$ denote the number of rows and columns of $\mathcal{M}$, respectively.
    \item[Phase 2] Differentiation. \\
    (a) Determine the set of differential equations $F_o$ if there exists $m_{ij} = 1$ for any derivative of $\mathbb{Y}$, and the set of algebraic equations $F_a$ if $m_{ij} = 0$ for all derivatives of $\mathbb{Y}$ in $f_i$, where $|F_o|=s$,  $|F_a|=n-s$.\\
    (b) Select the algebraic constraints $f_k (s+1 \leq k \leq n)$ from $F_a$ to differentiate $\upsilon_k$ such that $ord(y_j) \leq r_j$, which can be viewed as the homogeneous order. If it generates the new differential dependent variables,  then it requires to augment the row and column of variable pencil to denote $\mathcal{M'}$, update dependent variables set to $\mathbb{V'}$ and $\mathbb{V'}_{j}$. The terminated condition of algebraic differentiation is as follows:
      \begin{equation}\label{eqn:termcondition1}
      n + \sum_{k=s+1}^{n} \upsilon_k =|\mathbb{V'}| - |\mathbb{V'}_j| + 1. \ That \ is, \ row(\mathcal{M'}) = col(\mathcal{M'}) -|y_{j}^{[r_j]}|,
      \end{equation}
      where $\upsilon_1 = \upsilon_2 = \cdots=\upsilon_s =0$.\\
   (c)  Select some low order differential equations $f_k (1 \leq k \leq s)$  from $F_o$ to differentiate $\upsilon_{k}$ if (\ref{eqn:termcondition1}) fails such that $ord(f_k) \leq \max_{j=1}^{m} r_j$ with $y_j$, and augment the row and column of variable pencil to denote $\mathcal{M''}$, and update dependent variables set to $\mathbb{V''}$, $\mathbb{V''}_j$ and the order $r_j$ to $r'_j$. The terminated condition of differentiation is as follows:
     \begin{equation}\label{eqn:termcondition2}
      n + \sum_{k=1}^{n} \upsilon_k =|\mathbb{V''}| - |\mathbb{V''}_j| + 1. \ That \ is, \ row(\mathcal{M''}) = col(\mathcal{M''}) -|y_{j}^{[r'_j]}|.
      \end{equation}
\end{description}
\begin{remark}
We remark that the termination of index reduction procedure is required by the condition (\ref{eqn:termcondition1}) or (\ref{eqn:termcondition2}) because of the construction of a square elimination matrix. In particular, the procedure may have fully been degenerated into the problems of ODEs if (\ref{eqn:termcondition2}) fails in Phase 2(c). We can obtain the new set of ODEs from Phase 2(b) and (c), then refer to the algorithm of \cite{YZZ2012}.
\end{remark}
\begin{definition}
The number of differentiations specified by index reduction procedure gives a formula for the weak differentiation index of system of DAEs, denoted by $d_w = \max_{k=1}^{n} \upsilon_k$. Obviously, if no differentiation of the original system is index zero, ODEs may have the weak differentiation index more than zero.
\end{definition}
\begin{theorem}
Let $F=\{f_1, f_2, \cdots, f_n\} \in R[y_1, y_2, \cdots, y_m, y_1^{(1)}, \cdots, y_1^{(r_1)}, \cdots, y_m^{(1)}, \cdots, y_m^{(r_m)}]$, and the index reduction procedure satisfies the terminated condition (\ref{eqn:termcondition1}) or (\ref{eqn:termcondition2}). Then $\mathbb{\upsilon} = (\upsilon_1, \upsilon_2, \cdots, \upsilon_n)$ can be computed correctly as specified.
\end{theorem}
\begin{proof} 
The initialization in Phase 1, we can easily get the set of dependent variables $\mathbb{V}$ and construct the variable pencil $\mathcal{M}=(m_{ij})$ from $F$ and initialize $\upsilon_1= \upsilon_2=\cdots = \upsilon_n=0$.
According to Phase 2(a), we have 
\begin{equation}
\arraycolsep 0pt
\begin{array}{c}
    \{\underbrace{f_1, f_2, \cdots\cdots , f_s}_{\mathclap{ODEs\ (F_o)}} \  \underbrace{f_{s+1}, f_{s+2}, \cdots\cdots, f_n}_{\mathclap{algebraic \ equations\ (F_a)}} \}.
\end{array}
\end{equation}
To compute $\mathbb{\upsilon} = (\upsilon_1, \upsilon_2, \cdots, \upsilon_n)$ , two cases are considered:\\
Case (a): only differentiate $F_a = \{f_{s+1}, f_{s+2}, \cdots, f_{n}\}$ to satisfy the condition $(\ref{eqn:termcondition1})$ such that $ord(y_j) \leq r_j$ based on the homogeneous order, which can repeat the differentiation to obtain the differentiation times $\{\upsilon_{s+1}, \upsilon_{s+2}, \cdots, \upsilon_n\}$. In the general setting, since $|y_j^{[r_j]}|= r_j$, it is easy to get the terminated condition $(\ref{eqn:termcondition1})$. In particular, since $|y_j^{[r_j]}|<r_j$ for sparse case, it always generates the new differential dependent variables $\{y_1^{[k_1]}, y_2^{[k_2]}, \cdots, y_m^{[k_m]}\}$ with $k_j < r_j$, and requires to update the set of dependent variables $\mathbb{V'} = \mathbb{V}\cup y_1^{[k_1]}\cup y_2^{[k_2]} \cup \cdots \cup y_m^{[k_m]}$, and $\mathbb{F}= F \cup \{\delta f_{s+1}, \cdots, \delta^{\upsilon_{s+1}}f_{s+1}, \cdots, \delta {f_n}, \cdots, \delta^{\upsilon_{n}}f_{n}\}$. Consequently, we need to augment the row and column of variable pencil $\mathcal{M}$ to denote $\mathcal{M'}$. This concludes following:
\begin{equation*}
\ row(\mathcal{M'}) = col(\mathcal{M'}\setminus \{y_j, y_{j}^{[r_j]}\}) +1.
\end{equation*}
Case (b): following Case (a), if the condition $(\ref{eqn:termcondition1})$ fails, it needs to obtain the condition $(\ref{eqn:termcondition2})$. The problem can be transferred into the general $n \times m$ system of ODEs $F_o=F_o \cup \{\delta^{\upsilon_{s+1}}f_{s+1}, \cdots, \delta^{\upsilon_{n}}f_{n}\}$. In order to reduce the redundancy differentiation times, we only differentiate some low order ODEs from $F_o$ such that $ord(f_k) \leq r_j$ with $y_j$, which are $m_{ij} =0$ in $\mathcal{M'}$ for $\{ y_1^{(r_1)}, y_2^{(r_2)}, \cdots, y_m^{(r_m)}\}$. Furthermore, it may need to differentiate some general ODEs from $F_o$ to satisfy the condition $(\ref{eqn:termcondition2})$, which can repeat the differentiation to obtain the differentiation times $\{\upsilon_{1}, \upsilon_{2}, \cdots, \upsilon_n\}$. It always generates the new differential dependent variables $\{y_1^{[k'_1]}, y_2^{[k'_2]}, \cdots, y_m^{[k'_m]}\}$ and requires to update the set of dependent variables $\mathbb{V''} = \mathbb{V'}\cup y_1^{[k'_1]}\cup y_2^{[k'_2]} \cup \cdots \cup y_m^{[k'_m]}$, $\mathbb{F'}= \mathbb{F} \cup \{\delta f_{1}, \cdots, \delta^{\upsilon_{1}}f_{1}, \cdots, \delta {f_n}, \cdots,$ $\delta^{\upsilon_{n}}f_{n}\}$ and the order $r_j$ to $r'_j$. Consequently, we need to augment the row and column of variable pencil $\mathcal{M'}$ to denote $\mathcal{M''}$.  This concludes following:
\begin{equation*}
\ row(\mathcal{M''}) = col(\mathcal{M''}\setminus \{y_j, y_{j}^{[r'_j]}\}) +1.
\end{equation*}$\hfill \square$
\end{proof}
\begin{remark}
Yang et al \cite{YZZ2012} gave a formulation of differentiation times for differential elimination of ODEs. However, their method may lead to some redundant differentiation times in the practical applications, such as the constrained mechanical systems. Here, we propose a variable pencil technique to analyze the differentiation times of DAEs. It is able to make differentiation times as few as possible for differential algebraic elimination. It is also suitable for the differential elimination of ODEs with Dixon resultant formulation.
\end{remark}
We present a simple example to illustrate our index reduction procedure as follows:
\begin{example}\label{exam1}
This example is the linear, time-dependent index two DAE discussed in Gear \cite{G1988} as follows:
\begin{equation}
\begin{bmatrix}
  1 & \eta t \\
 0 & 0
  \end{bmatrix}
  \begin{pmatrix}
  \dot{y}_1\\
   \dot{y}_2\\
  \end{pmatrix}+\begin{bmatrix}
  0 & 1+\eta  \\
 1 & \eta t
  \end{bmatrix}
  \begin{pmatrix}
  {y}_1\\
  {y}_2\\
  \end{pmatrix}=\begin{pmatrix}
  {p}_1\\
  {p}_2\\
  \end{pmatrix},
\end{equation}
where the dependent variables $y_1, y_2$ with $t$ a scalar independent variable, ${p}_1$ and ${p}_2$ are the known forcing functions of $t$, and $\eta$ is a parameter. We can get the expanded form as follows:
\begin{equation}\label{exam1orig}
\left.
\begin{aligned}
0 &= f_1 = \dot{y}_1 + (1+\eta) y_2 + \eta t \dot{y}_2 - p_1, \\
0 &= f_2 =y_1 + \eta t y_2 - p_2.
\end{aligned} \ \ 
\right \}
\end{equation}
We can initialize the original system (\ref{exam1orig}) as follows:\\
(a) collect the set of dependent variable $\mathbb{V} = \{ \dot{y}_1, y_2, \dot{y}_2, y_1 \}$; (b) sort the set $\mathbb{V} = \{ y_1, \dot{y}_1, y_2, \dot{y}_2 \}$;\\
(c) construct the variable pencil
\begin{equation*}
\begin{array}{ll}
\mathcal{M} \Rightarrow \begin{blockarray}{ccccc}
& y_1 & \dot{y}_1 & y_2 & \dot{y}_2\\
\begin{block}{c[cccc]}
f_{1} & 0 & 1  & 1 &1\\
f_{2} & 1 & \fbox{0} & 1 & \fbox{0} \\
\end{block}
\end{blockarray}
\end{array}.
\end{equation*}
Obviously, we can get the $F_a$ and $F_o$ with $|F_a|=|F_o|=1$, and differentiate $f_2$ based on the homogeneous order as follows:
 \begin{equation*}
\begin{array}{ll}
\mathcal{M'} \Rightarrow \begin{blockarray}{ccccc}
& y_1 & \dot{y}_1 & y_2 & \dot{y}_2\\
\begin{block}{c[cccc]}
f_{1} & 0 & 1  & 1 &1\\
f_{2} & 1 & 0 & 1 & 0 \\
\end{block}
{\delta f_2}    & 0 & \fbox{1} & 1 &\fbox{1}  \\
\end{blockarray}
\end{array}.
\end{equation*}
Therefore, we have
\begin{equation}\label{exam1a}
0 = f_3 = \delta f_2 = \dot{y}_1 + \eta  y_2 + \eta t \dot{y}_2 - \dot{p}_2.
\end{equation}
For the differentiated equation $\delta f_2$ appended to the original system, the system of three equations $f_1$, $f_2$ and $f_3$ has four dependent variables $y_1, \dot{y}_1, y_2$, and $\dot{y}_2$. For eliminating the dependent variables $\{ y_1, \dot{y}_1\}$ or $\{ y_2, \dot{y}_2 \}$, the terminated condition of algebraic differentiation (\ref{eqn:termcondition1}) holds. Consequently, we can get the differentiation times $\mathbb{\upsilon} = (0, 1)$.
\end{example}
\begin{remark}
We only differentiate the equation $f_2$ once, i.e., $d_w =1 $, and mix the algebraic equations and ODEs to deal with uniformly. However, the existing methods need to differentiate $f_2$ twice until no algebraic equations appear by substitution, that is, the problem is index two \cite{G1988,RLW2001}.
\end{remark}

\section{Differential algebraic elimination}
\label{diffRes}
In this section, the definition of differential algebraic elimination of DAEs is introduced by using the generalized Dixon resultant formulation. Based on the index reduction algorithm in Section \ref{index}, we also present an algorithm for computing the differential algebraic resultant. Moreover, its basic properties are given.

\subsection{Definition of differential algebraic elimination}
The fundamental tool is based on the idea of algebraic Dixon resultant to create the differential algebraic elimination. Firstly, we construct the differential algebraic cancellation matrix, and then compute the entries of differential algebraic elimination matrix, determinant of which contains the differential algebraic resultant as a factor. That is, DAEs can be treated as polynomial system, and $y_j$ and its derivatives can be viewed as parameters, the other dependent variables and their derivatives as the purely algebraic variables are eliminated simultaneously. Therefore, we can obtain the single ODE with $y_j$ and its derivatives to directly apply the numerical method.

Let $\mathbb{F}=\{f_1, f_2, \cdots, f_n, \delta f_1, \cdots, \delta^{\upsilon_1}f_1, \cdots, \delta f_n, \cdots, \delta^{\upsilon_n}f_n\} \in R[y_1, y_2, \cdots, y_m, y_1^{(1)}, \cdots, y_1^{(r_1)}, \cdots,$ $ y_m^{(1)}, \cdots, y_m^{(r_m)}], \mathbb{\bar{Y}}=\{\bar{y}_1, \bar{y}_2, \cdots, \bar{y}_m, \bar{y}_1^{(1)}, \cdots, \bar{y}_1^{(r_1)}, \cdots, \bar{y}_m^{(1)}, \cdots, \bar{y}_m^{(r_m)}\}$, the system
\begin{equation}\label{eqn:origsys}
\{ f_1=0, f_2=0, \cdots, f_n=0\}
\end{equation}
has solution if and only if the system $\mathbb{F}$ has solutions for $\mathbb{\upsilon} = (\upsilon_1, \upsilon_2, \cdots, \upsilon_n)^T$. In order to define the differential algebraic elimination of (\ref{eqn:origsys}) it is necessary to find a weak differentiation index $\mathbb{\upsilon}$ for eliminating the $y_1, y_2, \cdots, y_{j-1}, y_{j+1}, $
$y_{j+2}, \cdots, y_m$ and their derivatives, such that $f_1, \cdots, f_n, \delta f_1, \cdots, \delta^{\upsilon_1}f_1, \cdots, \delta^{\upsilon_n}f_n$ are $(\upsilon_1 + 1) + (\upsilon_2 + 1) +\cdots + (\upsilon_n+1)$ polynomials in $\upsilon_1 + \upsilon_2 + \cdots + \upsilon_n + n-1$ variables.

By following the definition of Dixon resultant we have
\begin{definition}
Let $f_i$ be a differential polynomial in $R[y_1, y_2, \cdots, y_m, y_1^{(1)}, \cdots, y_1^{(r_1)}, \cdots, y_m^{(1)}, \cdots, y_m^{(r_m)}]$, $N = (\upsilon_1+1) +  (\upsilon_2+1) +\cdots + (\upsilon_n+1) - 1$ as mentioned above. The differential algebraic cancellation matrix $\mathcal{DC}_\mathbb{F}$ of $\mathbb{F}$ with $y_j$ and its derivatives is the $(N+1) \times (N+1)$ matrix as follows:
 \begin{equation*}
\begin{array}{c}
\mathcal{DC}_\mathbb{F} = \begin{bmatrix}
f_1(y_1, y_2, \cdots, y_m, y_1^{(1)}, \cdots, y_1^{(r_1)}, \cdots, y_m^{(r_m)}) & \cdots & f_{N+1}(y_1, y_2, \cdots, y_m, y_1^{(1)}, \cdots, y_1^{(r_1)}, \cdots, y_m^{(r_m)}) \\
f_1(\bar{y}_1, y_2, \cdots, y_m, y_1^{(1)}, \cdots, y_1^{(r_1)}, \cdots, y_m^{(r_m)}) & \cdots & f_{N+1}(\bar{y}_1, y_2, \cdots, y_m, y_1^{(1)}, \cdots, y_1^{(r_1)}, \cdots, y_m^{(r_m)}) \\
f_1(\bar{y}_1, \bar{y}_2,\cdots, y_m, y_1^{(1)}, \cdots, $ $y_1^{(r_1)}, \cdots, y_m^{(r_m)}) & \cdots & f_{N+1}(\bar{y}_1, \bar{y}_2, \cdots, y_m, y_1^{(1)}, \cdots, $ $ y_1^{(r_1)}, \cdots, y_m^{(r_m)}) \\
\vdots& \vdots & \vdots \\
f_1(\bar{y}_1, \bar{y}_2,  \cdots, \bar{y}_m, \bar{y}_1^{(1)}, \cdots, $ $\bar{y}_1^{(r_1)}, \cdots, \bar{y}_m^{(r_m)}) & \cdots & f_{N+1}(\bar{y}_1, \bar{y}_2,  \cdots, \bar{y}_m, \bar{y}_1^{(1)}, \cdots, $ $\bar{y}_1^{(r_1)}, \cdots, \bar{y}_m^{(r_m)})
\end{bmatrix},
\end{array}
\end{equation*}
where $\{y_j, y_j^{(1)}, \cdots, y_j^{(r_j)}\}$ as parameters do not replace by $\{\bar{y}_j, \bar{y}_j^{(1)}, \cdots, \bar{y}_j^{(r_j)}\}$ in $f_i (1\leq i \leq N+1)$. The differential algebraic elimination polynomial of $\mathbb{F}$ is denoted by $D\theta_\mathbb{F} \in R[\mathbb{Y}, \mathbb{\bar{Y}}]$,
\begin{equation}
D\theta_\mathbb{F} =\frac{det(\mathcal{DC}_\mathbb{F})}{\prod_{\substack{
i = 1\\
i \neq j}}^{m}(y_i-\bar{y}_i)(y_i^{(1)}-\bar{y}_i^{(1)})\cdots(y_i^{(r_i)}-\bar{y}_i^{(r_i)})},
\end{equation}
the row vector of differential algebraic elimination derived polynomials of $\mathbb{F}$ is denoted by $DP_\mathbb{F}$, and differential algebraic elimination matrix of $\mathbb{F}$ is denoted by $DA_\mathbb{F}$ as follows,
\begin{equation}
D\theta_\mathbb{F} =\\
\begin{pmatrix}
1\\
\vdots\\
(y_m^{(r_m)})^{Nd_{mr_m}-1}\\
\vdots\\
y_1^{d_1-1}\\
\vdots\\
\prod_{\substack{
i = 1\\
i \neq j}}^{m}\prod_{\mu_i=1}^{r_i}y_{i}^{id_i-1}(y_i^{(\mu_i)})^{(m+\sum_{k=1}^{i-1}r_{k}+\mu_i-1)d_{i\mu_i}-1}
\end{pmatrix}^T \cdot
DA_\mathbb{F} \cdot
\begin{pmatrix}
1\\
\vdots\\
(\bar{y}_{m}^{(r_m)})^{d_{mr_m}-1}\\
\vdots\\
\bar{y}_{1}^{Nd_1-1}\\
\vdots\\
\prod_{\substack{
i = 1\\
i \neq j}}^{m}\prod_{\mu_i=1}^{r_i}\bar{y}_{i}^{(N-i+1)d_i-1}(\bar{y}_i^{(\mu_i)})^{(N-m-\sum_{k=1}^{i-1}r_{k}-\mu_i+1)d_{i\mu_i}-1}
\end{pmatrix},
\end{equation}
where the rows and columns of $DA_\mathbb{F}$ are indexed ordering by $y_m^{(r_m)} \succ \cdots \succ y_m^{(1)} \succ \cdots \succ y_1^{(r_1)} \succ \cdots \succ y_1^{(1)} \succ y_m \succ \cdots \succ y_1$, $\bar{y}_m^{(r_m)} \succ \cdots \succ \bar{y}_m^{(1)} \succ \cdots \succ \bar{y}_1^{(r_1)} \succ \cdots \succ \bar{y}_1^{(1)} \succ \bar{y}_m \succ \cdots \succ \bar{y}_1$, respectively. The coefficient matrix $DA_\mathbb{F}$ is also a square matrix, determinant of which is called differential algebraic resultant, denoted by $DARes(y_j, y_{j}^{[r_j]})$.
\end{definition}
Here, we can write the $DA_\mathbb{F}$ in the following block structure notation:
\begin{equation}
\begin{array}{c}
DA_\mathbb{F} = \begin{bmatrix}
D_{0,0} & D_{0,1} &\cdots & D_{0, Nd_1-1}\\
D_{1,0} &D_{1,1} & \cdots & D_{1, Nd_1-1}\\
\vdots&\vdots &\ddots & \vdots \\
D_{d_1-1,0} & D_{d_1-1,1} & \cdots & D_{d_1-1,Nd_1-1}
\end{bmatrix},
\end{array}
\end{equation}
where each block $D_{ij}$ is of size $(N-r_j-1)!\prod_{\substack{
i = 2\\
i \neq j}}^{m}\prod_{\mu_i=1}^{r_i}d_id_{i\mu_i} \times (N-r_j-2)!\prod_{\substack{
i = 2\\
i \neq j}}^{m}\prod_{\mu_i=1}^{r_i}d_id_{i\mu_i}$. As the increasingly large scale system of DAEs, we can make use of its structure and block triangularization to decompose a problem into subproblems by permuting the rows and columns of a rectangular or square, unsymmetric matrix. For more details refer to \cite{PF1990}.

Following the properties of Dixon resultant we prove easily.
\begin{theorem}\label{thm:diffrescomp}
The differential algebraic elimination matrix $DA_\mathbb{F}$ is of size $N!\prod_{\substack{
i = 1\\
i \neq j}}^{m}\prod_{\mu_i=1}^{r_i}d_id_{i\mu_i} \times N!\prod_{\substack{
i = 1\\
i \neq j}}^{m}\prod_{\mu_i=1}^{r_i}d_id_{i\mu_i}$ at most, and the complexity of computing the entries of $DA_\mathbb{F}$ is $\mathcal{O}(d_{1}^2(N!\prod_{\substack{
i = 2\\
i \neq j}}^{m}$ $\prod_{\mu_i=1}^{r_i}d_id_{i\mu_i})^3)$ in the worst case, where $d_i$ and $d_{i\mu_i}$ are mentioned above.
\end{theorem}

\begin{theorem}\label{thm:extfactor}
Differential algebraic resultant can be expressed as a linear combination of  enlarged system of equations $\mathbb{F}$ with $y_j$,
\begin{equation}
DARes(y_j, y_{j}^{[r_j]})= \sum_{i=1}^{n}\sum_{\mu_i=0}^{\upsilon_i} \mathbb{K}_{i\mu_i} \delta^{\mu_i}f_i,
\end{equation}
where $\mathbb{K}_{i\mu_i}$ is a polynomial with respect to $\mathbb{Y}$ and can be deduced from $DP_\mathbb{F}$. Moreover, if $DARes(y_j, y_{j}^{[r_j]})$ is a reducible differential equation, it can also be proved that the extraneous factors mentioned above may include three parts which are taken from $DP_\mathbb{F}$, $DA_\mathbb{F}$ and the resulting resultant expression by substituting $DP_\mathbb{F}$, respectively.
\end{theorem}
\begin{remark}
From Theorem \ref{thm:extfactor}, we can remove the extraneous factors from differential algebraic resultant when the existing greatest common divisors in each row or column of differential algebraic elimination matrix. That is, the extraneous factors are the greatest common divisors in the algebraic cofactors of $DA_\mathbb{F}$.
\end{remark}
\begin{theorem}\label{thm:DAResnecessary}
Differential algebraic resultant is equal to zero that is a necessary condition for the existence of a common solution of system of DAEs.
\end{theorem}
\begin{corollary}
Let $y_1, y_2, \cdots, y_m$ be solutions of the system of DAEs (\ref{eqn:origsys}). Then $y_j$ satisfies the $DARes(y_j, y_{j}^{[r_j]})$. 
\end{corollary}

\subsection{Algorithm}
In this subsection, we have the following procedure for differential algebraic elimination.

\begin{description}
\item[Input:] DAEs system $F=\{ f_1=0, f_2=0, \cdots, f_n=0\}$, and dependent variables $\mathbb{Y}\setminus \{y_j, y_j^{[r_j]}\}$.
\item[Output:] a polynomial ODE only contains  $y_j$ and its derivatives.
\item[Step 1:] Count the number of DAEs and $\mathbb{Y}\setminus \{y_j, y_j^{[r_j]}\}$, denote $n $ and $m $ respectively, if $n$ is equal to $m$ plus 1, then goto Step 3.
\item[Step 2:] Call index reduction algorithm in Section \ref{index} by taking the $t$-derivative of $f_i$, update $n$ and $m$ such that $n=m+1$ by the enlarged system of equations $\mathbb{F}$ and new $\mathbb{Y}\setminus \{y_j, y_j^{[r_j]}\}$ , the collections are as follows,
\begin{equation}
 \label{eq:1}
\mathbb{F}  \gets \left\{ \begin{aligned}
f_1, & \ \delta f_1, & \cdots,& \ \delta^{\upsilon_1} f_1  \\
f_2, & \ \delta f_2, & \cdots,& \ \delta^{\upsilon_2} f_2  \\
&\ \vdots &&\ \vdots\\
 f_n, &\ \delta f_n, & \cdots,& \ \delta^{\upsilon_n} f_n
  \end{aligned} \right\} = 0, \ \ \ \ \ \ \  \ \ \ \ \ 
  \mathbb{Y} \gets \left\{ \begin{aligned}
y_1, &\ y_1^{(1)}, & \cdots,& \ y_1^{(r_1)}   \\
y_2, &\ y_2^{(1)}, & \cdots,& \ y_2^{(r_2)}   \\
&\ \vdots &&\ \vdots \\
 y_m, &\ y_m^{(1)}, & \cdots,& \ y_m^{(r_m)}   \\
  \end{aligned} \right\} \setminus \{y_j, y_j^{[r_j]}\}.
 \end{equation}

\item[Step 3:] Construct the differential algebraic cancellation matrix $\mathcal{DC}_\mathbb{F}$, obtain the entries of differential algebraic elimination matrix $DA_\mathbb{F}$, remove the greatest common divisors from each row or column of $DA_\mathbb{F}$, and then compute its determinant  $DARes(y_j, y_{j}^{[r_j]})$.
\item[Step 4:] Return  $DARes(y_j, y_{j}^{[r_j]})$.

\end{description}

\begin{theorem}\label{thm:darcom}
The above algorithm works correctly as specified and its complexity mainly contains index reduction algorithm and the computation of differential algebraic resultant.
\end{theorem}
\begin{proof}
Correctness of the algorithm follows from the Dixon elimination method. Regarding the dependent variables $\{y_j, y_j^{[r_j]}\}$ as parameters and the other ones as algebraic variables, we can treat the enlarged system of equations $\mathbb{F}$ as an algebraic system. As shown in \cite{YZZ2012}, a necessary condition for the existence of a common solution of algebraic differential equations is that the differential resultant is equal to zero. We can easily get the Theorem \ref{thm:DAResnecessary}.

From the description of algorithm, we observe that there are two major steps on time complexity. In Step 2, we can obtain the differentiation times $\sum_{k=1}^{n}\upsilon_k$. The problem is solved by homogeneous order rule, one that makes differentiation time as few as possible for reducing the number of enlarged system of equations $\mathbb{F}$. If there exists the $\mathbb{\upsilon} = (\upsilon_1, \upsilon_2, \cdots, \upsilon_n)^T$, which can be done in polynomial time. In Step 3, the complexity includes three parts: (a).  to obtain the entries of differential algebraic elimination matrix $DA_\mathbb{F}$ in Theorem \ref{thm:diffrescomp}, suppose $d_i=d_{i\mu_i}=d$, $r_i =r$, for each $i=1, 2, \cdots, m$,  it needs at most
\begin{equation*}
d_{1}^2(N!\prod_{\substack{
i = 2\\
i \neq j}}^{m}\prod_{\mu_i=1}^{r_i}d_id_{i\mu_i})^3 \leq (N!\prod_{\substack{
i = 1\\
i \neq j}}^{m}\prod_{\mu_i=1}^{r_i}d_id_{i\mu_i})^3 \leq (N!d^{(m-1)(m-1)r})^3 \leq  \mathcal{O}(N!^3d^{\mathcal{O}(m^2r)}),
\end{equation*}
which is the single exponential complexity; (b). calculate the greatest common divisors for each row or column of $DA_\mathbb{F}$ in the polynomial time; (c). compute its determinant  $DARes(y_j, y_{j}^{[r_j]})$ in polynomial time. Therefore, if there exists the differential algebraic resultant with single dependent variable and its derivative,  we can transform the system of DAEs to its equivalent ODEs in the single exponential complexity. $\hfill \square$
\end{proof}

\begin{remark}
From Lemma \ref{lem:KSY} and Theorem \ref{thm:darcom}, our algorithm is not a complete method. However, our algorithm is really effective and practical technique on numerous problems. It is well known that Dixon resultant elimination is that it can do one-step elimination of a block of unknowns from a system of polynomial equations like Macaulay's. Moreover, the size of Dixon matrix is much smaller than the size of Macaulay matrix. Though the entries of Dixon matrix are complicated in contrast to the entries in Macaulay matrix, the entries of which are either $0$ or coefficients of the monomials in the polynomial systems. Fortunately, we can easily apply the extended fast algorithm for constructing the Dixon matrix \cite{ZF2005}. In particular, for a fixed number $n$ of variables of a polynomial system, the construction cost of Dixon matrix is at most 
$\mathcal{O}(mvol(\mathcal{F})^3)$ arithmetic operations \cite{KS1996}, where $mvol(\mathcal{F})$ is the $n$-fold mixed volume. As shown in \cite{YZZ2012}, our algorithm is also appropriate to the system of ODEs. In many practical applications of DAEs, we can easily see that $d_i$ and $d_{i\mu_i}$ are very low degrees in $\mathbb{P}_{ik}$. 
\end{remark}
  
\begin{example}
Continue from Example \ref{exam1}, we can construct the differential algebraic elimination matrix with $y_1$ and $\dot{y}_1$ as follows:
\begin{equation*}
\begin{bmatrix}
\eta t p_1 - \eta t \dot{p}_2 + y_1 - p_2
  \end{bmatrix}_{1\times 1}.
   \end{equation*}
We can also eliminate $y_1$ and $\dot{y}_1$ by the same method simultaneously. Therefore, we get the following differential algebraic resultants:
\begin{equation*}
\begin{aligned}
DARes(y_1, \dot{y}_1) &= y_1 - p_2 + \eta t (p_1 - \dot{p}_2), \\
DARes(y_2, \dot{y}_2) &= y_2 - p_1 + \dot{p}_2.
\end{aligned}
\end{equation*}
These are the same as the results obtained in \cite{G1988}. 
\end{example}

\section{Examples}
\label{example}
In this section, we present some small examples in detail and compare the matrices size of differential resultants of two models with other methods. Examples \ref{exam2} and \ref{exam5} illuminate how to deal with the nonlinear and high index DAEs of constrained mechanical system. Example \ref{exam3}  uses a simple example to test our algorithm for the nonlinear and non-square system of DAEs. Example \ref{exam4} is a practical application for the linear electrical network problem. 

\subsection{Some examples in detail}
\begin{example}\label{exam2}
Consider the nonlinear DAEs for the simulation of the dynamics of multibody systems, which is a major application area. Here, we show the simple pendulum to illustrate many of the principles. The DAEs can be written
\begin{equation}
\left.
\begin{aligned}
0 &= f_1 = \ddot{y}_1 + y_1\lambda,\\
0 &= f_2 =\ddot{y}_2 + y_2\lambda - g, \\
0 &= f_3 =y_1^2 + y_2^2 - L^2, 
\end{aligned} \ \ \  \
\right \}
\end{equation}
where $g > 0, L > 0$ are constants. From (\ref{varpencil}) its variable pencil, labeled by equations and variables, is
 \begin{equation*}
\begin{array}{ll}
\mathcal{M} \Rightarrow \begin{blockarray}{cccccc}
& y_1 & \ddot{y}_1 & y_2 & \ddot{y}_2 & \lambda\\
\begin{block}{c[ccccc]}
f_{1} & 1 & 1  & 0 &0 & 1\\
f_{2} & 0 & 0 & 1 & 1 & 1 \\
f_{3} & 1 & \fbox{0} & 1 & \fbox{0} & 0 \\
\end{block}
\end{blockarray}
\end{array}.
\end{equation*}
Obviously, we can get the $F_a$ and $F_o$ with $|F_a|=1, |F_o|=2$, and differentiate $f_3$ based on the homogeneous order as follows:
 \begin{equation*}
\begin{array}{ll}
\mathcal{M'}  \Rightarrow \begin{blockarray}{cccccccc}
& y_1 &\dot{y}_1& \ddot{y}_1 & y_2 & \dot{y}_2 &\ddot{y}_2 & \lambda\\
\begin{block}{c[ccccccc]}
f_{1} & 1 & 0 & 1  & 0 & 0 & 0 & 1\\
f_{2} & 0 & 0 & 0  & 1 & 1 & 1 & 1\\
f_{3} & 1 & 0 & 0  & 1 & 0 & 0 & 0 \\
\end{block}
{\delta f_3}    & 1 & \fbox{1} & 0 &1 &\fbox{1} &0 &0 \\
{\delta^{2} f_3}    & 1 & 1& \fbox{1} & 1&1 &\fbox{1}&0  \\
\end{blockarray}
\end{array}.
\end{equation*}
Therefore, we have
\begin{equation}\label{exam2a}
\left.
\begin{aligned}
0 &= f_4 = \delta f_3 = 2y_1\dot{y}_1 + 2y_2\dot{y}_2 ,\\
0 &= f_5 = \delta^{2} f_3= 2y_1\ddot{y}_1 + 2y_2\ddot{y}_2 + 2\dot{y}_1^2 +2\dot{y}_2^2.
\end{aligned}\ \
\right \}
\end{equation}
For the differentiated equations $\delta f_3$ and $\delta^{2} f_3$ appended to the original system, the system of five equations $f_1, f_2, f_3, f_4$ and $f_5$ has seven dependent variables $y_1, \dot{y}_1, \ddot{y}_1, y_2, \dot{y}_2, \ddot{y}_2$, and $\lambda$. For eliminating the dependent variables $\{ y_1, \dot{y}_1, \ddot{y}_1\}$ or $\{ y_2, \dot{y}_2, \ddot{y}_2 \}$, the terminated condition of algebraic differentiation satisfies (\ref{eqn:termcondition1}). Consequently, we can get the differentiation times $\mathbb{\upsilon} = (0, 0, 2)$, that is, $d_w=2$.

We can construct the differential algebraic elimination matrix with $y_1, \dot{y}_1$ and $\ddot{y}_1$ as follows:
\begin{equation*}
\begin{bmatrix}
y_1\dot{y}_1& L^2-y_1^2& 0& 0& 0& 0& 0\\
0& 0& -y_1^2\ddot{y}_1-y_1\dot{y}_1^2& y_1\dot{y}_1& y_1^2-L^2& 0& 0\\
0& 0& y_1^2\dot{y}_1& L^2-y_1^2& 0& 0& 0 \\
1& 0& 0& 0& 0& 0& y_1^2\dot{y}_1 \\
0& -1& -g y_1& 0& 0& L^2-y_1^2& -y_1^2\ddot{y}_1-y_1\dot{y}_1^2\\
0& 0& -\ddot{y}_1& 0& 1& 0& -g y_1\\
0& 0& 0& 0& 0& -1& -\ddot{y}_1
 \end{bmatrix}.
   \end{equation*}
Obviously, we can also eliminate $y_1, \dot{y}_1$ and $\ddot{y}_1$ by the same method simultaneously. Therefore, we get the following differential algebraic resultants:
\begin{equation*}
\begin{aligned}
DARes(y_1, \dot{y}_1, \ddot{y}_1) &= (-2L^6y_1^2+L^8+y_1^4L^4)\ddot{y}_1^2+(-2L^4y_1^3+2L^6y_1)\dot{y}_1^2\ddot{y}_1+3g^2L^4y_1^4+L^4\dot{y}_1^4y_1^2+g^2y_1^8\\&-3g^2L^2y_1^6-L^6g^2y_1^2, \\
DARes(y_2, \dot{y}_2, \ddot{y}_2) &= (L^4-L^2 y_2^2)\ddot{y}_2+L^2\dot{y}_2^2y_2-g y_2^4+2gL^2y_2^2-g L^4. 
\end{aligned}
\end{equation*}
The remaining dependent variable $\lambda$ is determined by $y_1$ and $y_2$.
\end{example}

\begin{example}\label{exam3}
 The example is the nonlinear, non-square system of DAEs discussed in \cite{LGY2011} as follows:
\begin{equation}\begin{bmatrix}
  c_{10} &  0 & 0 &c_{13} \\
  c_{20} & 0 & c_{22} & 0 \\
   c_{30} & c_{31}& 0 & 0
  \end{bmatrix}
  \begin{pmatrix}
  1\\
  y_1y_2\\
   \dot{y}_1y_2\\
    \dot{y}_1\dot{y}_2
  \end{pmatrix}=\begin{bmatrix}
  0  \\
  0\\
 0
  \end{bmatrix},
\end{equation}
where the dependent variables $y_1$ and $y_2$, $c_{ij}(i=1, 2, 3, j=0, 1, 2, 3)$ are the known forcing functions of $t$. We can get the expanded form as follows:
\begin{equation}\label{exam3orig}
\left.
\begin{aligned}
0 &= f_1 =c_{10}+c_{13}\dot{y}_1\dot{y}_2,  \\
0 &= f_2 =c_{20}+c_{22}\dot{y}_1y_2,  \\
0 &= f_3 = c_{30}+c_{31}y_1y_2.
\end{aligned} \ \ \ \ 
\right \}
\end{equation}
We can initialize the original system (\ref{exam3orig}) as follows:\\
(a) collect the set of differential variable $\mathbb{V} = \{ \dot{y}_1, \dot{y}_2, y_2,  y_1 \}$; (b) sort the set $\mathbb{V} = \{ y_1, \dot{y}_1, y_2, \dot{y}_2 \}$;\\
(c) construct the variable pencil
\begin{equation*}
\begin{array}{ll}
\mathcal{M} \Rightarrow \begin{blockarray}{ccccc}
& y_1 & \dot{y}_1 & y_2 & \dot{y}_2\\
\begin{block}{c[cccc]}
f_{1} & 0 & 1  & 0 &1\\
f_{2} & 0 & 1 & 1 & 0 \\
f_{3} & 1 & \fbox{0} & 1 & \fbox{0} \\
\end{block}
\end{blockarray}
\end{array}.
\end{equation*}
Obviously, we can get the $F_a$ and $F_o$ with $|F_a|=1$ and $|F_o|=2$, and the system of three equations $f_1$, $f_2$ and $f_3$ has four dependent variables $y_1, \dot{y}_1, y_2$ and $\dot{y}_2$. For eliminating the dependent variables $\{ y_1, \dot{y}_1\}$ or $\{ y_2, \dot{y}_2 \}$, the terminated condition of algebraic differentiation satisfies (\ref{eqn:termcondition1}). Consequently, we can get the differentiation times $\mathbb{\upsilon} = (0, 0)$, that is, $d_w = 0$.

Here, we can construct the differential algebraic elimination matrix with $y_1$ and $\dot{y}_1$ as follows:
\begin{equation*}
\begin{bmatrix}
 c_{20}c_{31}y_1 - c_{22}c_{30}\dot{y}_1 
  \end{bmatrix}_{1\times 1}.
   \end{equation*}
We can eliminate $y_1$ and $\dot{y}_1$ by the same method simultaneously. Therefore, we get the following differential algebraic resultants:
\begin{equation*}
\begin{aligned}
DARes(y_1, \dot{y}_1) &=  c_{20}c_{31}y_1 - c_{22}c_{30}\dot{y}_1, \\
DARes(y_2, \dot{y}_2) &= -c_{10}c_{22}y_2 + c_{20}c_{13}\dot{y}_2. 
\end{aligned}
\end{equation*}
\end{example}

\begin{example}\label{exam4}
Consider a practical linear electrical network example, differential algebraic equations of index $1$ may have an arbitrarily high structural index from \cite{RMB2000} as follows:
\begin{equation}
\left.
\begin{aligned}
0 &= f_1 =  \dot{y}_2 + \dot{y}_3 + y_1 - a(t),\\
0 &= f_2 =  \dot{y}_2 + \dot{y}_3 + y_2 - b(t),\\
0 &= f_3 =  \dot{y}_4 + \dot{y}_5 + y_3 - c(t),\\
0 &= f_4 =  \dot{y}_4 + \dot{y}_5 + y_4 - d(t),\\
0 &= f_5 =   y_5 - e(t),\\
\end{aligned} \ \ \ \ 
\right\}
\end{equation}
where $a(t), b(t), c(t), d(t)$ and $e(t)$ are the known forcing functions of $t$. It is clear that $y_5$ is known, i.e., $y_5 = e(t)$. We can get the variable pencil as follows:
\begin{equation*}
\begin{array}{ll}
\mathcal{M} \Rightarrow \begin{blockarray}{cccccccccc}
& y_1 & y_2  & \dot{y}_2 & y_3 & \dot{y}_3  & y_4 & \dot{y}_4  & y_5 & \dot{y}_5 \\
\begin{block}{c[ccccccccc]}
f_{1} & 1 & 0 & 1  & 0 & 1 & 0 & 0 & 0 & 0\\
f_{2} & 0 & 1 & 1  & 0 & 1 & 0 & 0 & 0 & 0\\
f_{3} & 0 & 0 & 0  & 1 & 0 & 0 & 1 & 0 & 1\\
f_{4} & 0 & 0 & 0  & 0 & 0 & 1 & 1 & 0 & 1\\
f_{5} & 0 & 0 & 0  & 0 & 0 & 0 & 0 & 1 & \fbox{0}\\
\end{block}
\end{blockarray}
\end{array}.
\end{equation*}
Obviously, we can get the $F_a$ and $F_o$ with $|F_a|=1, |F_o|=4$, and differentiate $f_5$ based on the homogeneous order as follows:
 \begin{equation*}
\begin{array}{ll}
\mathcal{M'}  \Rightarrow \begin{blockarray}{cccccccccc}
& y_1 & y_2  & \dot{y}_2 & y_3 & \dot{y}_3  & y_4 & \dot{y}_4  & y_5 & \dot{y}_5 \\
\begin{block}{c[ccccccccc]}
f_{1} & 1 & 0 & 1  & 0 & 1 & 0 & 0 & 0 & 0\\
f_{2} & 0 & 1 & 1  & 0 & 1 & 0 & 0 & 0 & 0\\
f_{3} & 0 & 0 & 0  & 1 & 0 & 0 & 1 & 0 & 1\\
f_{4} & 0 & 0 & 0  & 0 & 0 & 1 & 1 & 0 & 1\\
f_{5} & 0 & 0 & 0  & 0 & 0 & 0 & 0 & 1 & 0\\
\end{block}
{\delta f_5}    & 0 & 0 & 0  & 0 & 0 & 0 & 0 &  0 &\fbox{1}\\
\end{blockarray}
\end{array}.
\end{equation*}
Therefore, we have
\begin{equation}\label{exam4a}
\begin{aligned}
0 &= f_6 = \delta f_5 = \dot{y}_5 - \dot{e}(t) .\\
\end{aligned}
\end{equation}
For the differentiated equations $\delta f_5$ appended to the original system, the system of six equations $f_1, f_2, f_3, f_4, f_5$ and $f_6$ has seven dependent variables $y_1, y_2, \dot{y}_2, y_3, \dot{y}_3, y_4$ and $\dot{y}_4$. For eliminating the dependent variables $\{ y_4, \dot{y}_4 \}$, $\{y_3, \dot{y}_3\}$ or $\{ y_2, \dot{y}_2 \}$, the terminated condition of algebraic differentiation satisfies (\ref{eqn:termcondition1}). Consequently, we can get the differentiation times $\mathbb{\upsilon} = (0, 0, 0, 0, 1)$. It only needs to differentiate $f_5$ once, that is, $d_w = 1$.
\end{example}

\begin{remark}
We easily compute the differential times $(0, 0, 1, 1, 2)$ of five equations $f_1, f_2, f_3, f_4, f_5$ in sequence by the $\Sigma$-method \cite{P2001}, that is, the structural index is 3. As for an increasingly large dimensions, the  $\Sigma$-method may perform an arbitrarily high differentiation times. However, our weak differentiation index is the same as the differentiation index. In general, it is suitable for the linear DAEs as follows,
\begin{equation}\label{eqn:gen}
A \dot{\mathbb{Y}} + B \mathbb{Y} = \mathbb{S},
\end{equation}
where $\mathbb{S}$ is the vector of $m$ sufficiently smooth forcing functions of $t$, $\mathbb{Y}$ is as mentioned above, $A$ and $B$ are $m \times m$ matrices, such as linear DAEs (\ref{eqn:gen}) with $m = 2 k + 1$, zero vector $\mathbb{S}$, and the identity matrix $B$, 
\begin{equation*}
\begin{array}{ll}
A = \begin{blockarray}{cccccccccc}
\begin{block}{[cccccccccc]}
 \cdot  & 1 & 1  &  &  &  &  &  & \\
              &  1 & 1  &  &  &  &  &  & \\
              &      & \cdot  & 1 & 1 &  &  &  & \\
              &      &              & 1 & 1 &  &  &  & \\
              &      &              &    &  \ddots &  &  &  & \\
               &      &              &   &  &\cdot &1  &1  &  & \\
                &      &              &  &  & & 1 & 1 &  & \\
                 &      &              &  &  & &  &  & \cdot & \\
\end{block}
\end{blockarray}
\end{array},
\end{equation*}
such that $A$ solely consists of $k$ blocks of form 
 $\bigl(\begin{smallmatrix}
1&1\\ 1&1
\end{smallmatrix} \bigr)$, the lower left element of each being on the main diagonal of $A$. This is the same result that the index is $1$ by using the Kronecker canonical form \cite{TI2008}. However, structural index algorithm \cite{P2001} needs to differentiate the last equation $k$ times, that is, the structural index is $k + 1$. In \cite{P1988}, Pantelides' algorithm needs to perform $k+1$ iterations before termination. Therefore, it leads to a large number of redundant differentiation times.
\end{remark}

\begin{example}\label{exam5}
We present a double pendulum model to demonstrate our index reduction technique in the dynamical systems. It is modeled by the motion in Cartesian coordinates, see Figure 1.
\begin{figure}[ht]\label{fig1}
\centering
\includegraphics[width=0.8\textwidth]{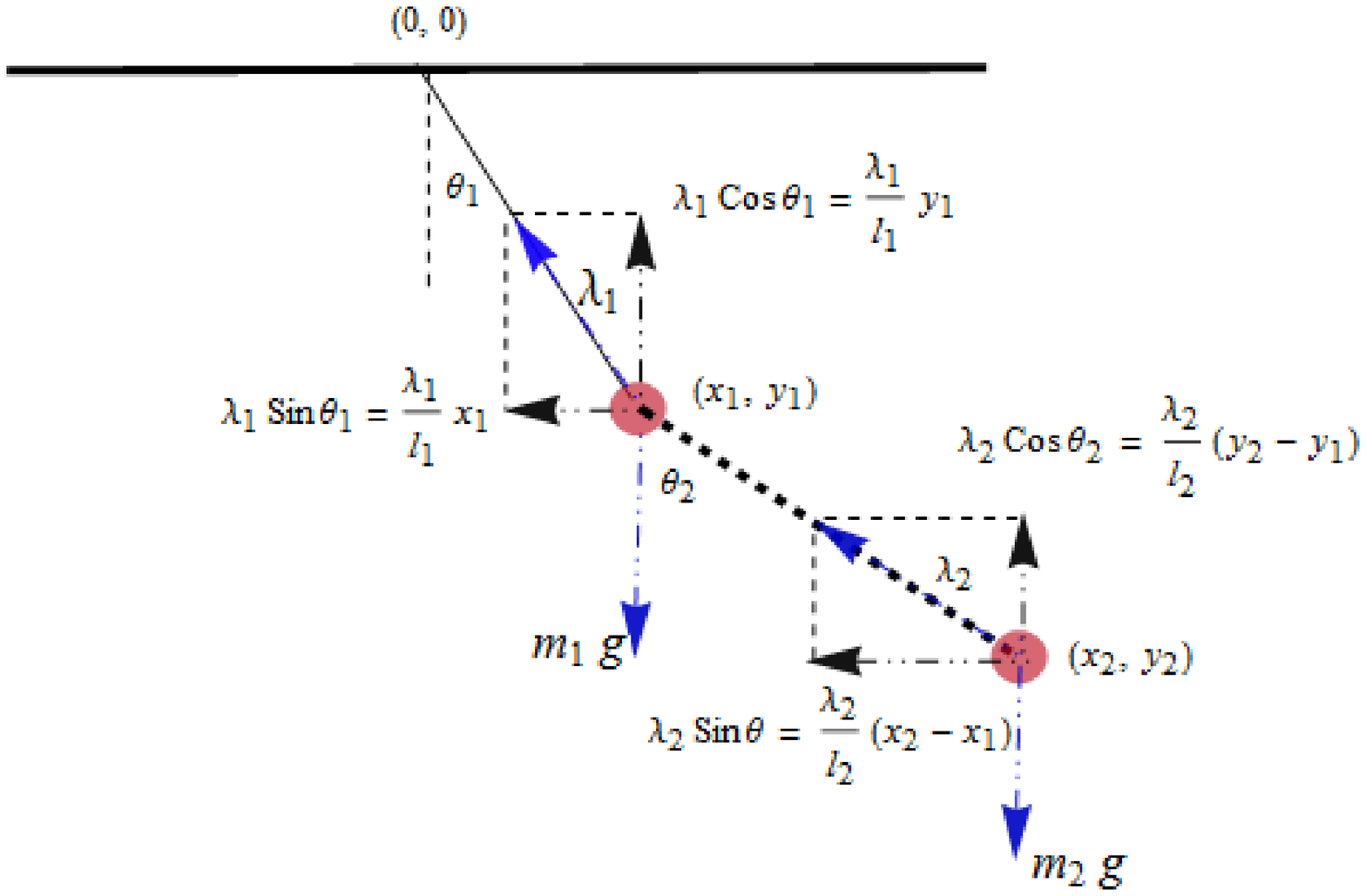}
\caption{Double Pendulum }
\end{figure}

We can derive the governing DAEs using Newton's second law of motion as follows,
\begin{equation}
\left.
\begin{aligned}
0 &= f_1 = m_1\ddot{x}_1 - \frac{\lambda_1}{l_1}x_1-\frac{\lambda_2}{l_2}(x_2 - x_1),\\
0 &= f_2 = m_1\ddot{y}_1 - \frac{\lambda_1}{l_1}y_1-\frac{\lambda_2}{l_2}(y_2 - y_1)-m_1g,\\
0 &= f_3 = m_2\ddot{x}_2 - \frac{\lambda_2}{l_2}(x_2 - x_1),\\
0 &= f_4 = m_2\ddot{y}_2 - \frac{\lambda_2}{l_2}(y_2 - y_1)-m_2g,\\
0 &= f_5 = x_1^2+y_1^2-l_1^2,\\
0 &= f_6 = (x_2-x_1)^2+(y_2-y_1)^2-l_2^2,\\
\end{aligned} \ \ \  \
\right \}
\end{equation}
where $g > 0, l_1>0, l_2 > 0, m_1>0$ and $m_2>0$ are constants, the dependent variables $x_1, x_2, y_1, y_2$ with $t$ a scalar independent variable, $\lambda_1, \lambda_2$ are the known forcing functions of $t$. 

From (\ref{varpencil}), we can construct its variable pencil $\mathcal{M}$, and easily to get the needed to differentiate $f_5$ and $f_6$ twice, respectively. Then we obtain the new variable pencil $\mathcal{M'}$ to determine the termination of differentiation, which holds the condition (\ref{eqn:termcondition1}). Finally, we can eliminate the dependent variables $\{x_1, \dot{x}_1, \ddot{x}_1\}, \{x_2, \dot{x}_2, \ddot{x}_2\}, \{y_1, \dot{y}_1, \ddot{y}_1\}$, or $\{y_2, \dot{y}_2, \ddot{y}_2\}$, where $\lambda_1, \lambda_2$ can be determined by $x_1, x_2, y_1$ and $y_2$. Therefore, we can get the differentiation times $\mathbb{\upsilon} = (0, 0, 0, 0, 2, 2)$, that is, $d_w = 2$.
\end{example}

\subsection{Some comparisons}
In this subsection, we also apply our algorithm to the system of ODEs and compare the matrix size of differential resultant with other methods. The algebraic manipulation of differential equations has gained importance in last years.  
\begin{example}\label{exam6}
Consider two nonlinear generic ordinary differential polynomials with order one and degree two from \cite{ZYG2014} as follows:
\begin{equation}
\left.
\begin{aligned}
0 &= f_1 = \dot{y}_1^2 + a_1 y_1 \dot{y}_1 + a_2 y_1^2 + a_3 \dot{y}_1 + a_4 y_1 + a_5, \\
0 &= f_2 = \dot{y}_1^2 + b_1 y_1 \dot{y}_1 + b_2 y_1^2 + b_3 \dot{y}_1 + b_4 y_1 + b_5, 
\end{aligned}\ \
\right \}
\end{equation}
where $a_i, b_i$ are differential constants, i.e., $\delta a_i = \delta b_i = 0 (i=1, 2, \cdots, 5)$. 
\end{example}

\begin{example}\label{exam7}
Consider the simplified version of a predator-prey model from \cite{CS2004} as follows:
\begin{equation}
\left.
\begin{aligned}
0 &= f_1 = a_2 y_1 + (a_1 + a_4  y_1)  y_2 + \dot{y}_2 + (a_3 + a_6 y_1) y_2^2 + a_5 y_2^3, \\
0 &= f_2 = \dot{y}_1 + (b_1 + b_3 y_1) y_2 + (b_2 +b_5 y_1) y_2^2 + b_4 y_2^3, 
\end{aligned} \ \
\right \}
\end{equation}
\end{example}
where $a_i, b_j (i=1, 2, \cdots, 6, j=1, 2, \cdots, 5)$ are the known forcing functions of $t$. 

\begin{table}[H]\label{tab1}
\begin{center}
\begin{tabular}{|c||c|c|c|}
 \hline \rotatebox{0}{Example}&\multicolumn{3}{c|} {\raisebox{.0ex}[0pt]{Matrix size} } \\
\cline{2-4}  & ZYG \cite{ZYG2014}&Rueda \cite{R2014}&Our algorithm\\
\hline\hline
\ref{exam6}& 36 $\times$ 36 & *& 9 $\times$ 9 \\ \hline 
\ref{exam7}&* & 13 $\times$ 13& 5 $\times$ 5 \\ \hline 
\end{tabular}
\caption{Matrix size for computing differential resultant}
\end{center}
\end{table}
Table 1 gives a comparison of the matrix size of differential resultant in Examples \ref{exam6} and \ref{exam7}, where '*' represents that the computation is not compared. From the Table 1, we have the following observations:

In two examples above, the matrix size of differential resultant via our algorithm is much smaller than two other methods. The smaller matrix leads to reduce more time for computing its symbolic matrix. It is consistent with the generalized Dixon resultant formulation. ZYG \cite{ZYG2014} is based on the idea of algebraic sparse resultant and Macaulay resultant for a class of the special ordinary differential polynomials. Rueda \cite{R2014} presents the differential elimination by differential specialization of Sylvester style matrices to focus on the sparsity with respect to the order of derivation.  In practice, Dixon's method is the most efficient technique to simultaneously eliminate several variables from a system of nonhomogeneous polynomial equations.  

\section{Conclusions}
In this paper, we propose a new index reduction for high index DAEs and establish a relationship between the generalized Dixon resultant formulation and system of DAEs  solving, which is defined as differential algebraic elimination. A significant problem in the differential algebraic elimination is to create methods to control the growth of differentiations. Our method can be applied to the mixed algebraic equations and differential equations to deal with simultaneously, and given a variable pencil technique to determine the termination of differentiation. From the algebraic geometry, it can be considered as the index reduction via symbolic computation. 

Our method can be also suitable for the system of ODEs and the high index nonlinear non-square system of DAEs, i.e., the number of dependent variable is not equal to the number of equations. The weak differentiation index is defined to unify the formulation of differentiation times for differential elimination of ODEs and differential algebraic elimination of DAEs. Moreover, a heuristics method is given for sifting the extraneous factors in differential algebraic resultants to remedy the drawback of factoring large polynomial system. Parallel computation can be used to speed up the computation of differential algebraic resultant of each dependent variable. 

However, the disadvantages of our method contain its limitation to polynomial coefficients and incomplete method because of the generalized Dixon elimination. Usually, for many practical relevant applications, the large scale system of ODEs/DAEs is also a challenge problem by the purely symbolic method; for instance, the full robot in the Modelica context \cite{F2015} has before symbolic simplification about 2391 equations and 254 dependent variables, which are reduced to 743 equations and 36 states that require a lot of index reduction going on. An obvious future work, is to attempt the block triangularization and sparsity considerations in constructing the differential algebraic elimination matrices. The sparseness is reflected in the quantity $l_i$ of $\mathbb{P}_{ik}$ in Section \ref{index}. Furthermore, symbolic-numeric differential algebraic elimination method is a very interesting work in the numerical algebraic geometry.

\section*{Acknowledgement}
This research was partly supported by China 973 Project NKBRPC-2011CB302402, the National Natural Science Foundation of
China (No. 61402537, 91118001), the West Light Foundation of Chinese Academy of Sciences, and the Open Project of Chongqing Key Laboratory of Automated Reasoning and Cognition (No. CARC2014004).

The first author is also grateful to Dr. Shizhong Zhao for his valuable discussions about removing the extraneous factors
from resultant computations.



\begin{thebibliography}{99}

\bibitem{BCP1996}
K. E. Brenan, S. L. Campbell, L. R. Petzold, Numerical Solution of Initial-Value Problems in Differential-Algebraic Equations, Second edition, Society for Industrial and Applied Mathematics, 1996.

\bibitem{C1993}
S. L. Campbell, Least squares completions for nonlinear differential algebraic equations, Numerische Mathematik 65(1993) 77--94.

\bibitem{CG1995}
S. L. Campbell, C. W. Gear, The index of general nonlinear DAEs, Numerische Mathematik 72(2)(1995) 173--196.

\bibitem{CK2003}
A. D. Chtcherba, D. Kapur, Exact resultants for corner-cut unmixed multivariate polynomial systems using the Dixon formulation, Journal of Symbolic Computation 36(3-4)(2003) 289--315.

\bibitem{CK2004}
A. D. Chtcherba, D. Kapur, Constructing Sylvester-type resultant matrices using the Dixon formulation, Journal of Symbolic Computation 38(1)(2004) 777--814.

\bibitem{CS2004}
A. C. Casal, A. S. Somolinos, Parametric excitation in a predator-prey model. In the first 60 years of Jean Mawhin, World Scientific, River Edge N.J., 2004, 41--54.

\bibitem{F1997}
G. Carr$\grave{a}$-Ferro, A Resultant Theory for the Systems of Two Ordinary Algebraic Differential Equations, Applicable Algebra in Engineering, Communication and Computing 8(1997) 539--560.

\bibitem{EH2009}
M. A. El-Khateb, H. S. Hussien, An optimization method for solving some differential algebraic equations, Communications in Nonlinear Science and Numerical Simulation 14(2009) 1970--1977.

\bibitem{F2015} P. Fritzson, Principles of Object-Oriented Modeling and Simulation with Modelica 3.3: A Cyber-Physical Approach, Second edition, Wiley-IEEE Press, 2015.

\bibitem{GLY2013}
X. S. Gao, W. Li, C.M. Yuan, Intersection theory in differential algebraic geometry: generic intersections and the differential Chow form, Transactions of the American Mathematical Society 365 (9) (2013) 4575--4632.

\bibitem{G1988}
C. Gear, Differential-Algebraic Equation Index Transformations, SIAM Journal on Scientific and Statistical Computing 9(1) (1988) 39--47.

\bibitem{HW1996}
E. Hairer, G. Wanner, Solving Ordinary Differential Equations II (second ed.), Springer-Verlag, Berlin, 1996.

\bibitem{KS1996}
D. Kapur, T. Saxena, Sparsity considerations in Dixon resultants, Proceedings
28th Annual ACM Symp. on Theory of Computation (STOC-28), Philadelphia, May 1996, 184--191.

\bibitem{KSY1994}
D. Kapur, T. Saxena, L. Yang, Algebraic and geometric reasoning using Dixon resultants, Proceedings of the International Symposium on Symbolic and Algebraic Computation, ACM, New York, 99--107, 1994.

\bibitem{K1973}
E. R. Kolchin, Differential Algebra and Algebraic Groups, Academic Press, London-New York, 1973.

\bibitem{KM2006}
P. Kunkel, V. Mehrmann, Differential-Algebraic Equations, Analysis and Numerical Solution, EMS Publishing House, Z$\ddot{u}$rich, Switzerland, 2006.

\bibitem{LGY2011}
W. Li, X. S. Gao, C. M. Yuan, Sparse Differential Resultant, Proceedings of the International Symposium on Symbolic and Algebraic Computation, ACM, New York, 225--232, 2011.

\bibitem{L2014}
C. S. Liu, A new sliding control strategy for nonlinear system solved by the Lie-group differential algebraic equation method, Communications in Nonlinear Science and Numerical Simulation 19(2014) 2012--2038.

\bibitem{M1992}
R. M$\ddot{a}$rz, Numerical methods for differentialÐalgebraic equations, Acta Numerica  (1)(1992) 141--198.

\bibitem{MS1993}
S. E. Mattsson, G. S$\ddot{o}$derlind, Index reduction in differential-algebraic equations using dummy derivatives, SIAM Journal on Scientific Computing 14(3)(1993) 677--692.

\bibitem{P1988} 
C. C. Pantelides, The consistent initialization of differential-algebraic systems, SIAM Journal on Scientific and Statistical Computing 9(2)(1988) 213--231.

\bibitem{PF1990} 
A. Pothen, C. J. Fan, Computing the block triangular form of a sparse matrix, ACM Transactions on Mathematical Software 16(4)(1990) 303--324.

\bibitem{P2001} 
J. D. Pryce, A simple structural analysis method for DAEs, BIT Numerical Mathematics 41(2) (2001) 364--394.

\bibitem{QWFR2013} 
X. L. Qin, W. Y. Wu, Y. Feng, et al., Structural analysis of high index DAE for process simulation, International Journal of Modeling, Simulation, and Scientific Computing 4(4)(2013) 1--16.

\bibitem{QSLF2014}
X. L. Qin, Z. Sun, T. Leng, et al., Computing the determinant of a matrix with polynomial entries by approximation, 2014. available at http://arxiv.org/pdf/1408.5879v2.pdf.

\bibitem{RLW2001}
G. J. Reid, P. Lin, A. D. Wittkopf, Differential Elimination-Completion Algorithms for DAE and PDAE, Studies in Applied Mathematics 106 (2001) 1--45.

\bibitem{RMB2000} G. Rei$\ss$ig, W. Martinson, P. I. Barton, Differential-algebraic equations of index 1 may have an arbitrarily high structural index, SIAM Journal on Scientific Computing 21(6)(2000) 1987--1990.

\bibitem{R1950}	
J. F. Ritt, Differential Algebra, Coll. Publ., Vol. 33, Amer. Math. Soc., New York, 1950.

\bibitem{R2013}
S. L. Rueda, Linear sparse differential resultant formulas, Linear Algebra and its Applications, 438 (11) (2013) 4296--4321.

\bibitem{R2014}
S. L. Rueda, Differential elimination by differential specialization of Sylvester style matrices, 2014. available at http://arxiv.org/pdf/1310.2081v2.pdf.

\bibitem{TI2008}
M. Takamatsu, S. Iwata, Index reduction for differential-algebraic equations by substitution method, Linear Algebra and its Applications 429(2008) 2268--2277.

\bibitem{WRI2009}
W. Y. Wu, G. Reid, S. Ilie, Implicit Riquier Bases for PDAE and their semi-discretizations, Journal of Symbolic Computation 44(7)(2009) 923--941.

\bibitem{YH1995}
L. Yang, X. R. Hou, Gather-and-Shift: a Symbolic Method for Solving Polynomial Systems, Proceedings of the First Asian Technology Conference on Mathematics, Association of Mathematics Educators, Singapore, 771--780, 1995.

\bibitem{YZZ2012}
L. Yang, Z. B. Zeng, W. N. Zhang, Differential elimination with Dixon resultants, Applied Mathematics and Computation 218 (2012) 10679--10690.

\bibitem{ZYG2014}
Z.Y. Zhang, C.M. Yuan, X. S. Gao, Matrix Formulae of Differential Resultant for First Order Generic Ordinary Differential Polynomials, Computer Mathematics, R. Feng et al. (eds.), Springer-Verlag Berlin Heidelberg, 479--503, 2014.

\bibitem{ZF2005}
S. Z. Zhao, H. G. Fu, An extended fast algorithm for constructing the Dixon resultant matrix, Science in China Series A: Mathematics 48(1)(2005) 131--143.

\bibitem{ZF2009}
S. Z. Zhao, H. G. Fu, Three kinds of extraneous factors in Dixon resultants, Science in China Series A: Mathematics 52(1)(2009) 160--172.

\bibitem{ZF2010}
S. Z. Zhao, H. G. Fu, Multivariate Sylvester resultant and extraneous factors (in Chinese), Science in China Series A: Mathematics 40(7)(2010) 649--660.



\end{thebibliography}
\end{document}